# RCNF: Real-time Collaborative Network Forensic Scheme for Evidence Analysis

Nour Moustafa, Jill Slay
School of Engineering and Information Technology
University of New South Wales at the Australian Defence Force Academy, Canberra, Australia
E-mail: nour.moustafa@unsw.edu.au, j.slay@.adfa.edu.au

**Abstract**

Network forensic techniques help in tracking different types of cyber attack by monitoring and inspecting network traffic. However, with the high speed and large sizes of current networks, and the sophisticated philosophy of attackers, in particular mimicking normal behaviour and/or erasing traces to avoid detection, investigating such crimes demands intelligent network forensic techniques. This paper suggests a real-time collaborative network Forensic scheme (RCNF) that can monitor and investigate cyber intrusions. The scheme includes three components of capturing and storing network data, selecting important network features using chi-square method and investigating abnormal events using a new technique called correntropy-variation. We provide a case study using the UNSW-NB15 dataset for evaluating the scheme, showing its high performance in terms of accuracy and false alarm rate compared with three recent state-of-the-art mechanisms.

1. Introduction

Due to the considerable increase of cyber attacks, network forensics isgrowing more sophisticated in methods used for investigating such attacks. For example, in May 2017, the WannaCry ransomware attack that targets Microsoft Windows operating systems infected more than 230,000 computer systems in about 150 countries, with the software requesting ransom payments in the cryptocurrency Bitcoin [15]. Identifying the origins of such attacks require thedevelopment of more network forensics investigation techniques t for analysing network traffic in order to identify the source of security policy abuse or information assurance violation [1] [3][4].

Although extracting network packets for forensic analysis is simple in theory, it requires an accurate inspection due to the current high speed and large size of networks and the collection of information from different heterogeneous sources [1] [5]. This accurate inspection needs an advanced feature selection method, which selects only relevant information, including attack patterns. Identifying the key features that capture important information is considered worthy of further intelligent analysis in order to aggregate network observations and investigate the attack evidence [16].

Multiple commercial and open source tools, including NIKSUN's NetDetector Suite, PyFlag and Xplico's tool, have been designed and applied to help network forensic examiners for carrying out the investigations of attacks [17]. These tools mainly depend on collecting flow information (i.e., IPs and ports) from network packets [1] [6]. However, such information is considered unreliable because of the mobile nature of appliances and the use of dynamic allocation of IP addresses [17]. Consequently, exploring the dependency of the flow information without analysing the transition between flows has become a challenge. This challenge is in investigating what has occurred in terms of the broader attack and who was actually invlved in order to build an effective network forensic framework [17].

In this study, we propose a Real-time Collaborative Network Forensic scheme (RCNF) that can monitor and track the origins of cyber attacks. The scheme involves three components. Firstly, capturing network data using a tcpdump sniffer tool [21], and then storing it in a MySQL database [22] in order to be much easier for analysing and aggregating network data using a suggested aggregator module. After this, important features include potential characteristics of abnormal activities are selected using the chi-square method [18] [25]. Finally, we develop a correntropy variation technique that can specify a Risk Level (RL) for normal and attack observations as evidence analysis, as discussed in Section 4. The proposed RCNF and its modules are evaluated on the UNSW-NB15 dataset [14] as it has a wide range of contemporary authentic normal and abnormal observations.

The main contributions in this paper are elaborated as follows. First, we propose a RCNF Scheme for investigating attack activities on large-scale networks. Second, we develop two new components in the scheme of aggregating network flows to reduce irrelevant observations and the Chi-Square feature selection method for reducing irrelevant features, in addition to a correntropy-variation technique for defining attack vectors and specifying its risk level.

The rest of the paper is organised as follows. The related work and existing network forensic techniques and frameworks are discussed in Section 2. Section 3 discusses the network forensic process while Section 4 provides the details of the RCNF scheme and its components. The results and discussions are provided in Section 5. Finally, the paper summary and future directions are presented in Section 6.

## 2. Network forensic techniques and related work

Several network forensic methods have been utilized to vulnerabilities [1] [10] [12]. To begin with, the logging method is used for recording network data in a database to inspect attack evidence. Different attributes should be stored, for example, flow identifiers source/destination IP addresses and ports, and some statistical information about packets, such as a packet size and interval-packet length. Different algorithms, such as apriori, hypothesis testing, protocol analysis and immune, are used to track attack activities from the logged files [1] [10]. Secondly, a packet-marking method is applied to mark network packets at different routers while sending flows from a sender to its receiver [12]. Machine learning and heuristic approaches are also widely used for modelling and investigating attack events throughout networks [1] [13]. These approaches help to detect different normal and suspicious instances in the training phases and then validates the approaches' correctness for recognising suspicious instances in the testing phase [14].

Multiple network forensic frameworks have been proposed [1] [2] [3] [4] [5]. Firstly, a traceback-based framework is used to identify the origin of network packets, which are used for investigating attack paths of Distributed Denial of Service (DDoS) and IP spoofing attacks [2]. Multiple trace-back mechanisms should be incorporated in order to defend against these attacks from large-scale networks. For example, Wang et al. [3] developed a topology assisted deterministic packet marking technique based on an IP traceback for tracking DoS and DDoS attacks. Cheng et al. [4] suggested a cloud-based traceback architecture in order to tackle the access control challenge in cloud systems. The goal of the study is to prevent normal users from requesting traceback information for malicious intentions.

Secondly, a converged network-based framework defines digital evidence in VoIP communication [1] [5]. Voice packet is breached by attackers during voice communication which changes a normal voice packet to a suspicious one. For instance, Ibrahim et al. [5] designed a VoIP evidence model for investigating attacks in VoIP communication by making a hypothesis based on information collected. In [6], the authors used some existing network forensics techniques to model network vulnerability and network evidence graph. In it, network vulnerability evidence and reasoning techniques were used for reconstructing malicious scenarios and then backtracking the network packets to get the original evidence. Thirdly, an attack graph-based framework discovers and visualises all possible attack paths throughout a network by analysing computer and network systems [1]. A probabilistic method [8] was proposed based on Bayesian inference for designing evidence graphs. This method can address false positive rates and inspect evidence by computing the posterior probabilities.

The distributive based framework is also proposed for investigating cyber crimes in order to handle the scalability problem by distributing network forensic servers and data agent systems [1]. In [9], a network forensic architecture is suggested. This architecture comprises five components: collection and indexing, database management, an analysis component, analysis communication component and the database for collecting and analysing abnormal patterns. Finally, a network intrusion detection framework is used for monitoring and protecting malicious activities. Network Forensics based on intrusion detection systems executes static and dynamic inspection for network abnormal data [10]. Wang et al. [11] proposed a hybrid attack detection and forensic technique in machine-to-machine networks. In addition, this technique was designed for recognising DDoS attacks in a distributed anti-honeypot- based forensics architecture.

Although existing frameworks can investigate attack activities to some extent, the high speed and large sizes of current networks cause a big challenge of extracting important information that can be used for investigating the origin of attacks, and a methodology of investigating abnormal activities with a confidence level that reveals to what extent the risks of those activities. Moreover, existing network forensics techniques often consume high computational resources while investigating large-scale distributed networks without aggregating relevant flows that include an attack. These challenges are the motivation of our study and are addressed by designing an aggregation component that can collect only relevant information, as explained in Section 4. Then, we propose a correntropy-variation mechanism for investigating attacks by specifying a risk level for each network observation.

## 3. Network forensic process

The key functions of a network forensics technique are to log attackers' behaviour and provide a forensic technique in order to inspect the data logged. Therefore, the information of attackers and the invasion process can be easily discovered [20]. A typical network forensics mechanism includes the basic steps for investigating and discovering the paths between the victims and attackers, shown in Figure 1. The network forensics mechanism analyses the capturing data and provides the analysis results, revealing the evidence of capturing anomalous observations based on the new correntropy-variation technique as discussed below.

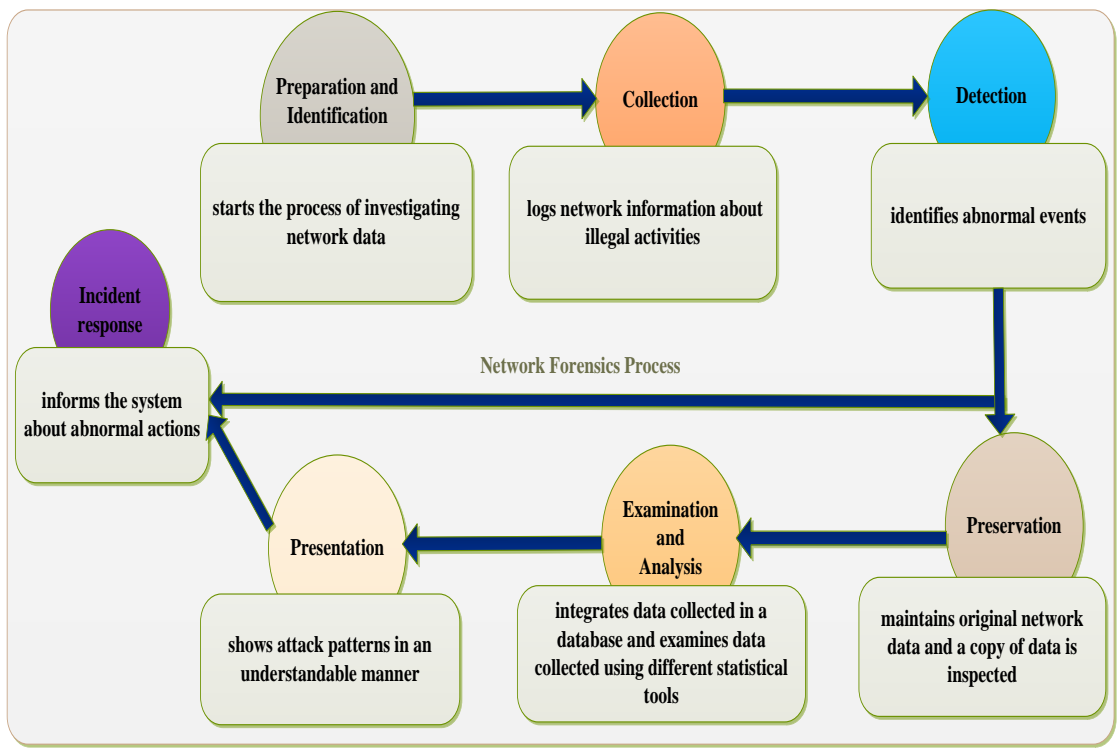

Figure 1: Steps of network forensics techniques

## 4. Real-time Collaborative Network Forensic scheme

The proposed Real-time Collaborative Network Forensic (RCNF) scheme comprises three key modules, as presented in Figure 2. Capturing and storing network data is the first stage to sniff network packets and log them into a database to make it much easier while investigating attack patterns. Then, selecting important features is the second stage to remove any redundant information that affects recognising attack activities. Finally, investigating attack events is a crucial phase of defining abnormal events and their origins, as detailed in the following three subsections.

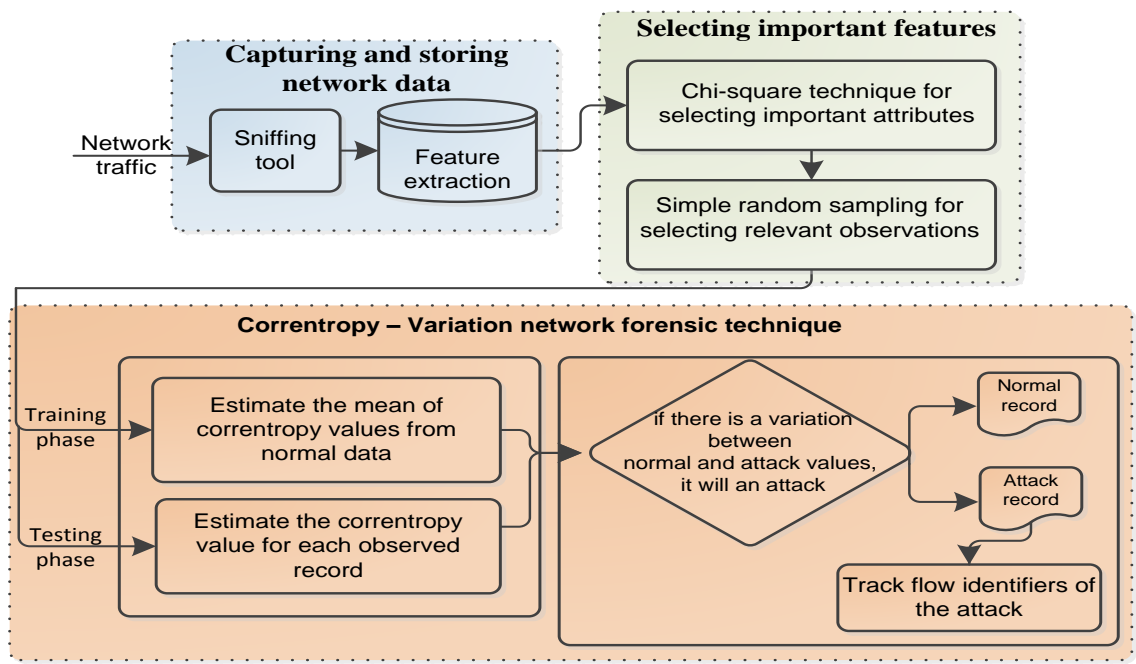

Figure 2: proposed RCNF architecture

### 4.1 Capturing and storing network data

The large number of flows of current networks demands an aggregation method starting from capturing packets to storing them in a database for summarising network activities and building the proposed RFCN. A tcpdump tool [21] is utilised for capturing the raw packets from the network interfaces as it reduces dropping packets in the production systems. Then, a number of features are generated from the packets using Bro and Argus tools, as in the UNSW-NB15 dataset. Network traffic has to be sniffed at the checkpoints, in particular, ingress routers in order to collect only relevant flows based on their source and destination IP addresses and protocols. This leads to decrease the computational processing time and assists to investigate the origins of cyber adversaries using an advanced network forensic technique.

The target of designing a collaborative RCNF can be achieved via collecting flows from all destination points across a network and storing those flows in a MySQL database that can be shared with different network forensics instances installed on the network. The flows are recorded using the technology of MySQL Cluster CGE [22], which can handle large-scale data in real-time processing.

The flows are aggregated via the MySQL's functions [23] for grouping the data using more than one attribute of the flow identifiers. This addresses the drawbacks of the existing tools, such as Netflow and sFlows that can accumulate only one feature each time. The paths of attacks are easily investigated when the flows between the source and destination IP addresses are counted and tracked. For example, DDoS attacks send large numbers of pings to a specific target to disrupt its resources; therefore, if these pings are monitored and numbered, the origins of the attackers can be easily identified using the network forensic technique discussed in subsection 4.3. For tracking the non-stationary properties of flow identifiers, we apply the '*count*' functions to determine all possible combinations of these flows, as follows.

- *Select COUNT(\*) as flows, srcip, dstip from network_data group by srcip, dstip;*
- *Select COUNT(\*) as flows, srcip, srcport from network_data group by srcip, srcport;*
- *Select COUNT(\*) as flows, dstip, dsport from network_data group by dstip, dsport, srcport;*

In the above queries, *flows* denote the number of flows, which occurred between any two attributes, *srcip* refers to the source IP address, *dstip* refers to the destination IP address, *srcport* refers to the source port, *dsport* is the destination port, and *proto* refers to the protocols. Every query retrieves the number of flows which takes place amongst the features.

A real-time and collaborative network forensics technique can be executed when the flows collected do not include duplicated ones or missing values. Consequently, we use the Simple Random Sampling (SRS) approach which selects a sample arbitrarily where no flows are included more than once within a sample size. All subsets of the examples are given equal probabilities. Moreover, any given pair of values has a similar probability of selection as the other pairs, reducing data bias and simplifying data analysis in which n samples are picked out of N examples [24].

**4.2 Selecting important features**

Besides selecting only relevant flows, the important features in the flows should be adopted to design the RFCN. We apply the chi-square feature selection method ($x^2$) [18] [25] due to its simplicity of implementation at real-time. Statistically speaking, the $x^2$ is used for measuring the occurrences of two independent variables associated with their class label, and then the highest ranked variables are selected as important features using equation (1).

$$x^2 = \sum_{i=1}^{y} \sum_{j=1}^{c} \frac{(O_{i,j} - E_{i,j})^2}{E_{i,j}} \tag{1}$$

where $x^2$ refers to the chi-square of Independence, $O_{i,j}$ denotes the observed value of two variables and $E_{i,j}$ is the mean of two variables.

### 4.3 Correntropy-variation technique for network forensics

The correntropy-variation technique is a combination of correntropy [26] measure for estimating the similarities between normal and attack instances, and a variation threshold for discovering attacks. The correntropy is a nonlinear similarity function that reveals the relationships between normal and abnormal observations, while the variation estimates how far the abnormal instances from the normal ones.

A correntropy of two random variables ($f_1$ and $f_2$) is estimated by

$$V_\sigma(f_1, f_2) = E\left[k_\sigma(f_1 - f_2)\right] \quad (2)$$

where $E[.]$ refers to the mean of features, $k_\sigma(.)$ is the Gaussian kernel function and $\sigma$ is the kernel size computed via

$$K_\sigma(.) = \frac{1}{\sqrt{2\pi}\sigma} \exp(-\frac{(.)^2}{2\sigma^2}) \quad (3)$$

The joint probability density function ($P_{F1,F2}(f_1, f_2)$) is unidentified, whilst a finite number of observations ($\{f_i, f_j\}_{i,j=1}^{M}$) is known. Consequently, the correntropy is measured via

$$V_{M,\sigma}(A, B) = \frac{1}{M} \sum_{i,j=1}^{M} K_\sigma(f_i - f_j) \quad (4)$$

To apply the correntropy for multivariate network data, as provided in equation (5), we calculate it for both normal and malicious observations as

$$I_{1:N} = \begin{bmatrix} f_{11} & f_{12} & \ddots \\ f_{21} & f_{22} & f_{ij} \end{bmatrix}, Y_{1:N} = \begin{bmatrix} c_1 \\ c_i \end{bmatrix} \quad (5)$$

such that $I$ is the observations of network data, $Y$ is the class label ($c$) of each observation, $N$ is the number of observations and $F$ is the number of features.

The mean of correntropy values of normal vectors ($corpy^{normal}$) is computed using equation 6 in the training phase. In the testing phase, the correntropy value ($corpy^{test}$) is estimated for each record based on equations 4 and 5. We design a baseline between the $\mu(corpy^{normal})$ and each $corpy^{test}$ using the standard deviation measure $(\sigma)$, which estimates the amount of variation between the mean of normal correntropy values and each correntropy of testing records. If the variation between the two values is greater than or equal $(2\sigma)$, the testing vector is considered as an attack, as given in equation 7. This is because such a vector is so far from the dispersion of normal correntropy values and is difficult to fit it within the same distribution of normal data. We called this threshold a Risk Level $(RL)$ that can identify all attack observations with low false alarm rates. The RL is scaled in a range of [0, 1] in order to exactly specify to what extent the abnormal activates deviate from normal ones.

$$\mu(corpy^{normal}) = \frac{1}{N}(corpy^{normal}) \quad (6)$$

$$RL = \begin{cases} \mu(corpy^{normal}) - (corpy^{test}) \geq 2\sigma & attack \\ else & normal \end{cases} \quad (7)$$

The origins of attack instances can be easily tracked via correlating their flow identifiers with their estimated RL. This way will help to define the risk level of those instances. If the RL value equals one, this means that type of attacks constitutes the highest risk to an organisation as it sends many flows to a specific destination such as events of DDoS attacks. But. if the RL value equals zero, this indicates this type of attacks makes the lowest risk to that organisation. For example, Table 1 lists some flow identifiers from the UNSW-NB15 dataset with estimated RL values. We observe that the abnormal records have higher RL (more than 0.5) than normal activities (less than 0.5). Ultimately, the proposed network forensic technique can define attack activities and their risk level, helping network administrators to track and report bad events that try to penetrate their network.

Table 1: Selected vectors with Risk Level (RL)

| srcip | sport | dstip | dsport | proto | label | RL |
|---|---|---|---|---|---|---|
| 149.171.126.14 | 179 | 175.45.176.3 | 33159 | tcp | 0 | 0.23 |
| 175.45.176.1 | 15982 | 149.171.126.14 | 5060 | udp | 0 | 0.11 |
| 175.45.176.3 | 63888 | 149.171.126.14 | 179 | tcp | 0 | 0.25 |
| 175.45.176.2 | 7434 | 149.171.126.16 | 80 | tcp | 1 | 0.83 |
| 175.45.176.0 | 15558 | 149.171.126.13 | 179 | tcp | 1 | 0.72 |

## 5. Empirical results and discussions

### 5.1 Dataset used and evaluation metrics

In order to assess the performance of our proposed scheme, we used the UNSW-NB15 dataset because it comprises a large collection of contemporary legitimate and anomalous vectors. The size of its network packets is approximately 100 Gigabytes extracted 2,540,044 feature vectors and was recorded in four CSV files. Each vector contains 47 features and the class label. It involves ten classes, one normal and nine types of security events and malware, namely Analysis, DoS, Exploits, Fuzzers for suspicious activity, Generic, Reconnaissance, Backdoors, Shellcode and Worms.

The evaluation criteria of accuracy and False Alarm Rate (FAR) are applied to measure the performance of the proposed scheme for identifying and tracking attack vectors. These criteria are described as follows.

- **Accuracy-** is the percentage of legitimate and suspicious vectors that are correctly identified, that is,

$$Accuracy = \frac{TP + TN}{TP + TN + FP + FN} \qquad (8)$$

- **FAR**- is the percentage of normal and malicious vectors that are incorrectly classified, that is,

$$FAR = \frac{FP + FN}{TP + TN + FP + FN} \qquad (9)$$

## 5.2 Pre-processing and selecting feature and observation stages

The proposed mechanisms were developed using the 'R language' on Windows 7 OS with 16 GB RAM and an i7 CPU processor. To carry out the experiments, we chosen arbitrary samples from the UNSW-NB15 dataset with several sample sizes between 100,000 and 300,000 for selecting the important features using the chi-square method and investigating attack activities by the correntropy-variation technique. In Table 2, eight important features are adopted using the chi-square technique based on their high weight.

Table 2: Features selected for investigating attacks

| Weight | Feature name | Feature description |
|--------|--------------|---------------------|
| 0.592  | sbytes       | source to destination bytes |
| 0.558  | swin         | source TCP window advertisement |
| 0.552  | dttl         | destination to source time to live |
| 0.551  | stcpb        | source TCP sequence number |
| 0.550  | dtcpb        | destination TCP sequence number |
| 0.549  | dwin         | destination TCP window advertisement |
| 0.513  | smean        | mean of the flow packet size transmitted by the source |
| 0.489  | sload        | source bits per second |

The feature vectors and flow identifiers (i.e., source IP (srcip), source port (sport), destination IP (dstip), destination source (dsport) and protocol types (proto)) are selected using the SRS technique in order to remove repeated instances or missing values, improving the overall performance of the RCNF scheme. As demonstrated in Table 1, an example of select five vectors from the USNW-NB15 dataset was designed to show how the risk level is computed. Then, these levels are connected with their flow identifiers for analysis the evidence of attack activities. The SRS and chi-square techniques ensure selecting the relevant observations and features that reflect the patterns of legitimate and suspicious instances while running the network forensic technique.

**5.3 Network forensic evaluation**

It clear that the correntropy makes a clear difference between the legitimate and attack feature vectors, as it estimates the nonlinear similarities between these vectors. As shown in Figure 3, the correntropy values of 2000 normal samples are obviously different from attack ones. As a result, the different attack types can be considerably identified and investigated their paths using the five flow identifiers of source/destination IP addresses and protocol types associated with their risk level, as listed in Table 1.

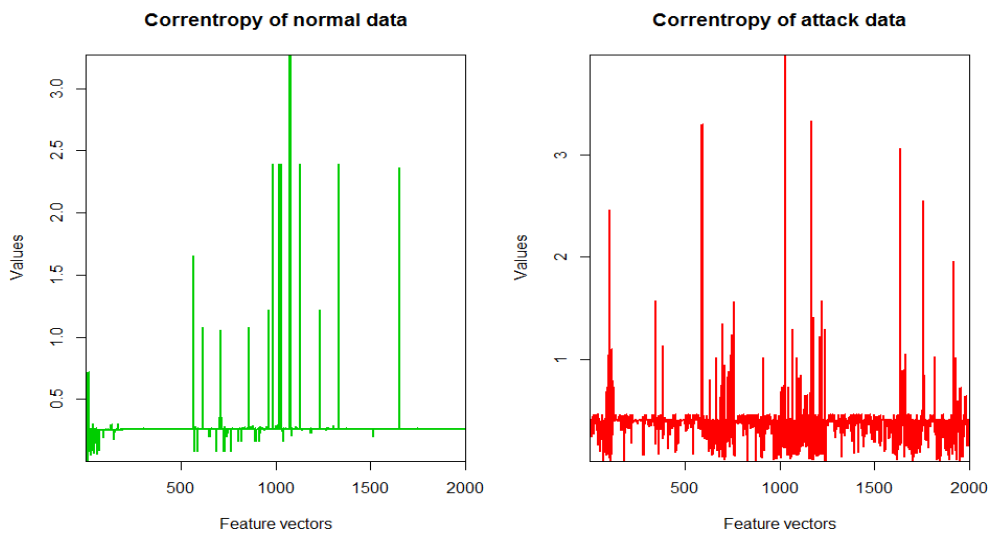

Figure 3: correntropy of some normal and attack samples

The performance of the technique is evaluated in terms of the overall accuracy and FAR on the feature selected in Table 1. Moreover, the ROC curve which shows the relationship between the accuracy and FAR with different three sample sizes is shown in Figure 3. The overall accuracy improved from 94.31% to 95.98%, whilst the overall FAR reduced from 5.69% to 4.02 % with increasing the sample sizes of data from 100,000 to 300,000.

Table 3: Performance of correntropy-variation technique

| Sample size | Accuracy | FAR |
|---|---|---|
| 100,000 | 94.31% | 5.69 % |
| 200,000 | 95.72% | 4.28 % |
| 300,000 | 95.98% | 4.02 % |

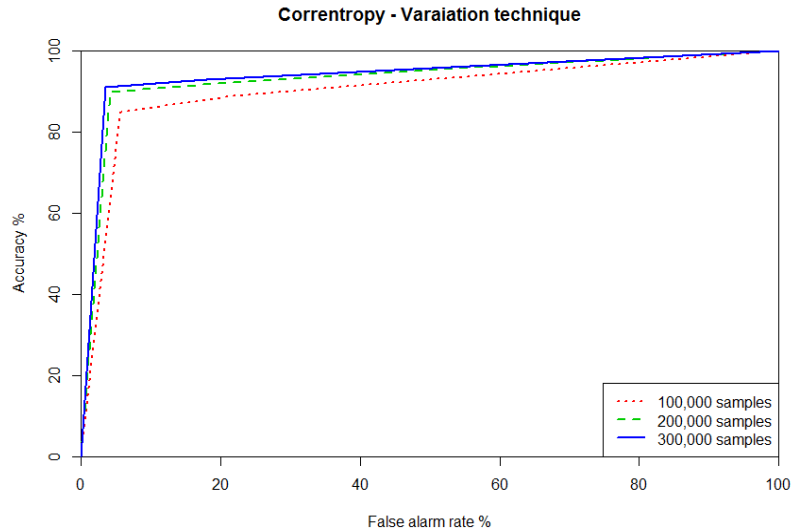

Figure 4: ROC curves of correntropy-variation technique for three sample sizes

The proposed mechanism can effectively recognise different attack vectors in the dataset, as declared in Table 4. The accuracy of detecting normal vectors increases from 92.12% to 93.29% while increasing the sample size from 100,000 to 300,000. Likewise, the accuracy of recognising malicious vectors rise up progressively from an average of 45.82% to an average of 97.55%

Table 4: Comparison of vector accuracy on three sample sizes

| | Sample size | | |
|---|---|---|---|
| Vector types | 100,000 | 200,000 | 300,000 |
| Normal | 92.12% | 93.16% | 93.29% |
| Exploits | 76.47% | 77.82% | 77.19% |
| Backdoor | 54.42% | 71.23% | 72.42% |
| Shellcode | 65.76% | 66.48% | 65.98% |
| Worms | 45.82% | 45.92% | 48.87% |
| DoS | 95.71% | 95.13% | 97.55% |
| Analysis | 88.26% | 89.45% | 90.22% |
| Fuzzers | 64.33% | 65.23% | 66.28% |
| Reconnaissance | 58.38% | 59.24% | 60.32% |
| Generic | 83.56% | 87.52% | 88.87% |

The Correntropy-Variation network forensic (CV-NF) technique is compared with three state-of-the art techniques, namely Filter-based Support Vector Machine (FSVM) [27], Multivariate Correlation Analysis (MCA) [28] and Artificial Immune System (AIS) [29] using the UNSW-NB15 dataset. The results revealed that the CV-NF technique outperforms the other mechanisms in terms of accuracy and FAR, as depicted in Figure 5.

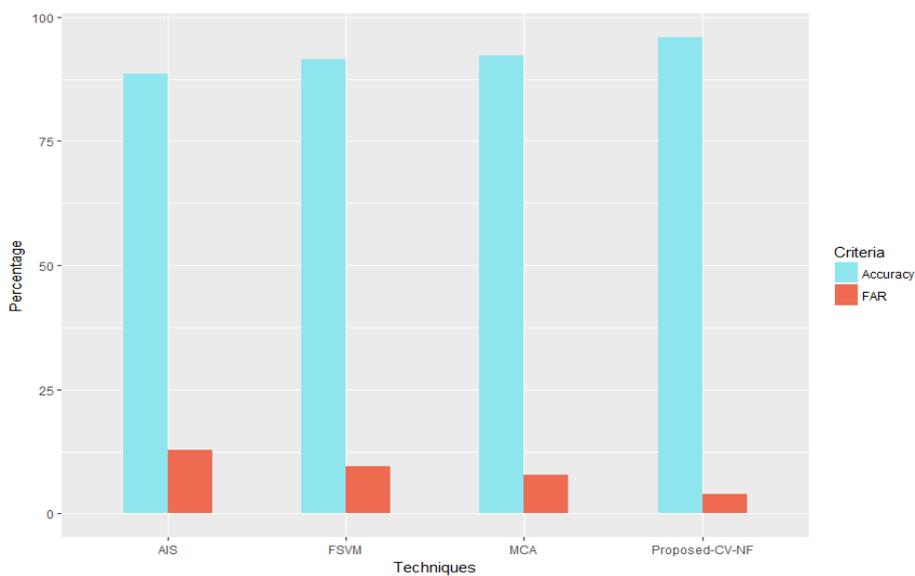

Figure 5: Comparison of performance of four techniques

The reason for outnumbering the CV-NF technique is that it developed based on estimating the correntropy values for normal and testing observations, and then considering any observation varies than 2σ of normal measures as attacks. The FSVM and AIS mechanism depend on training legitimate and malicious observations with large numbers of vectors to be correctly trained and validated, whereas the MCA technique relies on only computing correlations between attributes with the Gaussian mixture model to identify the DoS attacks, which sometimes cannot precisely identify the boundaries between legitimate and suspicious mixture models [30].

## 6. Conclusion and future directions

This paper discussed a new real-time collaborative network Forensic scheme (RCNF) for monitoring and defining the origins of cyber-attacks. The scheme involves three key steps: capturing and storing network data, selecting important network features and investigating abnormal activities. The important observations and features are selected using the chi-square and SRS methods, respectively, whilst investigating and identifying suspicious events are developed using a correntropy variation technique in order to define high-risk levels as attack evidence. The analysis evidence of attacks by correlating risk levels estimated with their flow identifiers. The scheme has been evaluated using the UNSW-NB15 dataset, and the results revealed its efficiency and effectiveness, in terms of accuracy and error rates, compared with three existing mechanisms. In future, we will apply the proposed scheme in cloud and fog computing systems due to their prevalence in our era.